\documentstyle[aps,prd,preprint]{revtex}

\tightenlines

\newcommand{\beq}{\begin{equation}}
\newcommand{\eeq}{\end{equation}}
\def\beqa{\begin{eqnarray}}
\def\eeqa{\end{eqnarray}}

\def\nn{\nonumber}
\def\q{\quad}

\begin{document}

\title{Gravitating global defects:
the gravitational field and compactification}

\author{Itsaso Olasagasti}

\address{Institute of Cosmology,
Physics Department, Tufts University, Medford, MA 02155, USA}

\maketitle

\begin{abstract}
We give a prescription to add the gravitational field of
a global topological defect to a solution of
Einstein's equations in an arbitrary number of dimensions. 
We only demand that the original solution
has a $O(n)$ invariance with $n\geq 3$. We will see that
the general effect of a global
defect is to introduce a deficit solid angle.
We also show how the same kind of scalar field configurations can be
used for
spontaneous compactification of $n$ extra dimensions on an $n$-sphere.

\end{abstract}

\section{Introduction}

Global defects arise in theories in which a global symmetry is broken
and the
vacuum manifold is nontrivial \cite{defects}. An example where this
kind
of
topological defect can arise is given by a theory with
$n$ scalar fields and a potential
which has a vacuum manifold with the topology of a
$(n-1)-$sphere. This is the case we will consider here.

Global defects
were first considered  in four spacetime dimensions, but more
recently they have also been studied in the context of the brane-world
scenario. In four spacetime dimensions,
when $n=2$, the defects are
global strings \cite{ruth}
and when $n=3$, global monopoles \cite{v}.
In the brane-world scenarios one models our 'visible' universe
as a $3$-brane living in a higher-dimensional spacetime with
$n$ extra dimensions. In this setup
one can imagine the $3-$brane being the
core of a global defect. Such a possibility has been investigated in a
number of papers. The domain wall case, a single extra dimension,
was studied long ago in Ref.\cite{ruba}. Explicit geometries
in higher dimensions are more recent.
The case with $n=2$, two extra dimensions,
has been considered in \cite{ck} and \cite{ruth2}.
In \cite{ov} solutions to Einstein's field equations (EFE's)
with $n>2$, three and higher extra dimensions were given, 
in \cite{sha} the localization of the
graviton in some of those solutions was discussed, while
in Ref.\cite{oda} Oda addressed
the issue of localization of various fields.

In the present paper we want to study further the gravitational
effects of global defects for $n\geq 3$. We will show how to construct
new
solutions to Einstein's equations from known ones. We will see that
the net effect of the addition of a global charge, away from the core,
is
to introduce a deficit solid angle, just as it was shown for the global
monopole, $n=3$, in four dimensions by Barriola and Vilenkin \cite{v}.

We will also show that higher-dimensional solutions with
geometry $\cal{M}$${}^{q}\times S^{n-1}$
can be constructed, where $\cal{M}$$^{q}$ is a solution to
$q$-dimensional Einstein gravity.

The plan of the paper is as follows. In the next section
we introduce the class of global defect
that we are going to study.
In section III we
show how to add the gravitational field of the global defect
to a known solution of Einstein's equations.
In section IV we use the global defect
for compactification on a $(n-1)$-sphere. Conclusions are summarized in
section V.

\section{global defects and the $\boldmath{\sigma}$-model
approximation}

In this section we present the class of models that we have
in mind.

To keep the discussion as general as possible, in this section we
consider spacetime to be $d$-dimensional with a metric of the form
\beq
ds^2=g_{A B} dx^A dx^B=g_{i j}(x^i) dx^{i}dx^{j}+B^2(x^{i})d\Omega^2_{n-1}
\eeq
where $d\Omega^2_{n-1}$ is the metric on a $(n-1)$-sphere of unit radius which
is parametrized by the $n-1$ angular coordinates $y^{\alpha}$.

The models that we are interested in include
a multiplet of
$n$ scalar fields $\phi^ a$, $a=1...n$
with a potential $V(\phi^a\phi^a)$ that has a minimum  on the
$(n-1)$-sphere
$\phi^a \phi^a=\eta^2$. We further assume that
$V(\phi^a\phi^a=\eta^2)=0$, since
a nonzero value can be absorbed into the cosmological constant term
in Einstein's equations.

It is well known that if the mapping
of a $(n-1)$-sphere
in configuration space in the vacuum manifold
is non-trivial, continuity requires the fields
to take values outside the vacuum manifold in some region surrounding
the origin. This region is called the core of the defect.
At a distance from the core, where the fields are in the vacuum
manifold,
the lagrangian is that for $n$ scalar fields subject to
the constraint $\phi^a \phi^a=\eta^2$. Outside the core the system is
thus well
approximated by a nonlinear $\sigma-$model, and the
equations for the fields can be written as
\beq
\nabla^2 \phi^a+\eta^{-2}(\partial_A \phi^b \partial^A \phi^b) \phi^a=0
\eeq

We shall consider the simplest non-trivial
{\em ansatz}, the familiar hedgehog configuration, in which
the fields depend only on the angular variables $y^{\alpha}$
through the relation
$\vec{\phi}(\vec{r})=\eta \hat{r}$, where $\hat{r}$ is the unit vector
on the
$(n-1)$-sphere. 
%when embedded in an $n+1$ euclidean space.
It is not hard to check that this configuration does indeed satisfy
the field equations.

To solve Einstein's equations we need the expression for the energy
momentum tensor. For the hedgehog configuration it is
\beq
T^i_{ j}=-{1 \over 2}  (n-1)
{\eta^2 \over B^2(x^i)} \;\delta^i_{j},
\q\q
T^{\alpha} _{\beta}=-{1 \over 2}(n-3)
{\eta^2\over B^2(x^i)} \;\delta^{\alpha} _{\beta}.
\label{emt}
\eeq

It is apparent from this expressions that if $B^2(x^i)$ goes to zero
at some point, the energy-momentum tensor is singular.
This signals the presence of the defect core where
continuity
requires that fields go to zero, which in turn
removes the singularity of the energy-momentum tensor.
We see how the pressence
of the defect is related to the fact that
the metric function $B^2(x^i)$ vanishes at some point.
Accordingly, if $B^2(x^i)$ does not vanish at any point, the
fields can stay in the vacuum manifold over the whole space. If this is
the case, no defect is present and the $\sigma$-model is exactly
applicable
over the entire space. An example is any
geometry of the form $\cal{M}$${}^{(d-n+1)}\times S^{n-1}$ in which
$n-1$ dimensions are compactified on a sphere. We shall
consider both cases in the following sections.

\section{Adding a global defect}
In this section we are going to show that just as the effect of local
strings is to
introduce a deficit angle in the plane transverse to the string, the
effect
of a global defect in the hedgehog configuration is to introduce a
deficit solid angle.
We will also give a prescription to construct solutions to Einstein's
equations with global topological
charges in the $\sigma$-model approximation described above.

We start by considering a solution to Einstein's equations in 
$d=q+n-1$
dimensions in the following form:
\beqa
ds_0^2&=&g_{ij}(x^i) dx^i dx^j+A^2(x^i) d\Omega^2_{n-1}
\label{ds0} \\
d\Omega^2_{n-1}&=& \hat{g}_{\alpha \beta}(y^{\beta}) dy^{\alpha}
dy^{\beta}
\eeqa
with $i,j=1..q$ and $\alpha,\beta=1...n-1$. The $\{y^{\alpha}\}$ are
the angular coordinates in $d\Omega^2_{n-1}$.
This is a general example of a warped geometry with warp factor
$A^2(x^i)$.
For such geometries
the Ricci tensor splits as
\beqa
R^0_{ij}&=&\tilde{R}_{ij}-(n-1)\;{A;_{ij}\over A}  \label{sp1}\\
R^0_{\alpha \beta}&=&\hat{R}_{\alpha \beta}
-\hat{g}_{\alpha \beta}\left[
A \tilde{\Box}A+(n-2)\tilde{\nabla} A  \tilde{\nabla} A
\right] \label{sp2}
\eeqa
The $0$ superscripts refer to
quantities for the metric (\ref{ds0}).
Tildes stand for quantities
calculated with the $q-$dimensional metric $g_{ij}(x^i) dx^i dx^j$
and the hats refer to quantities for 
$d\Omega^2_{n-1}$.

We assume that the above is a solution to EFE's
with a
stress-energy tensor $T^0_{A B}$,

\beq
R^0_{A B}-{1 \over 2}\;g^0_{A B} R^0= \kappa^2 T^0_{A B}
\eeq

We will now modify the original metric by
multiplying the warp factor by a constant factor
$(1-\Delta)$. The line element thus obtained is

\beq
ds^2=g_{ij}(x^i) dx^i dx^j+(1-\Delta)A^2(x^i) d\Omega^2_{n-1}
\eeq

This will be a solution to EFE's for a different energy-momentum
tensor $T_{AB}=\kappa^{-2}G_{AB}$ which we want to relate to
$T^0_{AB}$.

We can use equations (\ref{sp1}) and (\ref{sp2}) to
write the Ricci tensor of the new metric $R_{AB}$
in terms of $R^0_{AB}$

\beqa
R_{ij}&=&R^0_{ij} \\
R_{\alpha \beta}&=&\Delta \hat{R}_{\alpha \beta}+(1-\Delta)
R^0_{\alpha \beta}\\
R&=& R^0+{\Delta \over 1-\Delta} {\hat{R}\over A^2}
\eeqa

With these relations we construct the Einstein tensor from which we can
directly read off $T_{A B}$. We find that

\beqa
T^i_j&=&(T^0){}^i_j-{1\over 2}\;\delta^i_j{\Delta \over \kappa^2
(1-\Delta)}
   {\hat{R}\over A^2}, \\
T^{\alpha}_{\beta} &=& (T^0) {}^{\alpha}_{\beta}+
{\Delta \over \kappa^2 (1-\Delta)}
{\hat{G}^{\alpha}_{ \beta}\over A^ 2}.
\eeqa

Since the hats refer to $d\Omega^2_{n-1}$, we have 
$\hat{R}=(n-1)(n-2)$ and
$\hat{G}_{\alpha \beta}=-(n-3)(n-2)\hat{g}_{\alpha \beta}/2$. 
Putting these expressions into the equations above, we see
that the effect
of introducing the $(1-\Delta)$ factor in the metric (\ref{ds0})
corresponds to the addition of
matter with energy-momentum tensor given by

\beqa
T^i_j&=&-{1 \over 2} \delta^i_j
{\Delta (n-2)\over \kappa^2 (1-\Delta)} {n-1 \over A^2}
   \label{add1}\\
T^{\alpha}_{\beta}&=&-{1 \over 2} \delta^{\alpha}_{\beta}
{\Delta (n-2)\over \kappa^2 (1-\Delta)}
{n-3 \over A^2}
\label{add2}
\eeqa

It is clear that this is equivalent to (\ref{emt})
with $B^2(x^i)=(1-\Delta)A^2(x^i)$
when we choose $\Delta$ to have the
value:
\beq
\Delta= {\kappa^2 \eta^2 \over n-2}
\eeq

We thus have a prescription to add the gravitational field of a
topological
global charge to a known solution with $O(n)$ invariance: we just
have to
multiply the $d\Omega^2_{n-1}$ piece of the line element by the factor
$1-(\kappa^2\eta^2/(n-2))$, which we can interpret as
a deficit solid angle.

We now apply the prescription described above to some examples.

For the usual four-dimensional spacetime
the result presented above
is applicable to every isotropic solution to Einstein's equations.
Straightforward examples are the FRW
cosmological models and the Scharwzchild 
and Reissner-Nordstrom black hole solutions.

The metric for the global monopole in a general FRW spacetime
will look like
\beq
ds^2=-dt^2+a^2(t)\left\{ {dr^2 \over 1+kr^2}
+(1-\kappa^2\eta^2 )r^2 d\Omega^2_2 \right\}
\eeq

For the Schwarschild and Reissner-Nordstrom solutions
the metric, after the inclusion of the global
charge, has the form

\beq
ds^2=-f(r)dt^2+{dr^2\over f(r)}+(1-\kappa^2\eta^2 )r^2 d\Omega^2_2
\eeq

The first thing we must note is that, due to the deficit solid angle,
the metric is no longer asymptotically flat and so the no-hair theorem
for
black holes does not
apply in this case. The metric for the Schwarzschild solution with
a global topological charge was first derived by Barriola and Vilenkin
in \cite{v}. As they point out, such a topologically charged black hole
could
arise after the collision of a global monopole and a black hole.
An analysis along the lines of \cite{l}, where the solution
describing a black hole with a local cosmic string passing through it
was given,
shows that the area law for the black hole entropy still holds.

A global charge can also be added to black $p-$branes \cite{Duff}
\beq
ds^2=e^{2A}(-e^{2f} dt^2 +d\vec{x}^2)+e^{2B}\left[e^{-2f} dr^2
+r^2 \left(1-{\kappa^2\eta^2\over n-2}\right) d\Omega_{n-1}^2 \right].
\eeq

See Ref. \cite{Duff} for details about the
metric functions $A$, $B$ and $f$
as well as the bulk gauge and scalar fields also present.
Likewise we can add the far field of a global defect in any brane in the
same way, as long as the metric induced on the brane has the required
$O(n)$ symmetry.

\section{compactification}
In this section we show that it is possible to construct new compactified
solutions
from lower dimensional ones by adding appropriate scalar fields.
In this case we start with a general $q-$dimensional metric $ds_0^2$
and
consider the higher dimensional metric

\beqa
ds^2&=&ds_0^2+C^2d\Omega^2_{n-1}, \\
ds_0^2&=&g_{ij}(x^i) dx^i dx^j,
\label{ds0p} \\
C^2 d\Omega^2_{n-1}&\equiv&C^2 \hat{g}_{\alpha \beta} dy^{\alpha}
dy^{\beta}
\equiv {g}_{\alpha \beta} dy^{\alpha} dy^{\beta}, \\
i,j&=&1...q, \q
\alpha,\beta= 1...n-1,
\eeqa
where $C^2$ is a constant and $\hat{g}_{\alpha \beta}$ the metric
components on the unit $(n-1)$-sphere.

We can now compute the Einstein tensor for the extended solution.
Let's take
$G^0_{ij}=\kappa^2 T^0_{ij}$. Then we find that
\beqa
G_{ij}&=&G^0_{ij}-{1\over 2}\;g_{ij}{\hat{R}\over C^2} \nn \\
      &=& \kappa^2 T^0_{ij}-{1\over 2}\;g_{ij}
      {(n-1)(n-2)\over C^2} \label{efe1}\\
G_{\alpha \beta} &=&\hat{G}_{\alpha \beta}+{\kappa^2\over q-2}
g_{\alpha \beta} T^0 \nn \\
       &=&{\kappa^2\over q-2} g_{\alpha
\beta} T^0-{1\over 2}\;{(n-2)(n-3)\over C^2} {g}_{\alpha \beta}
\label{efe2}
\eeqa
where $T^0\equiv g^{ij} T^0_{ij}$.

As we did in the previous section, we will show that the extra
contribution can be obtained
from a multiplet of $n$ scalar fields
in the hedgehog configuration.

Let's consider first the case in which the lower-dimensional metric
is a solution to EFE's with a cosmological constant $\Lambda^0$:
$T^0_{ij}=-\Lambda^0 g_{ij}$.
We assume that in the
higher-dimensional case the energy-momentum tensor has contributions
from a cosmological constant $\Lambda$, not necessarily the same as
$\Lambda^0$, and the scalar fields. The contribution from the scalar 
fields is
given by expression (\ref{emt}) with $B^2(x^i)=C^2$ so that
\beqa
T_{ij}&\equiv& -\Lambda g_{ij}
      -{1 \over 2}  (n-1)
{\eta^2 \over C^2} \;g_{ij},       \label{t1}\\
T_{\alpha \beta}&\equiv&-\Lambda g_{\alpha \beta}
 -{1 \over 2}(n-3)
{\eta^2\over C^2} \;g_{\alpha \beta}. \label{t2}
\eeqa

From equations (\ref{efe1}),
(\ref{efe2}), (\ref{t1}) and (\ref{t2}), we see that
the higher-dimensional
EFE's reduce to
\beqa
-\Lambda^0 -{1\over 2}{(n-1)(n-2)\over C^2}
&=& -\Lambda- {n-1 \over 2} \;\;
{\kappa^2  \eta^2 \over C^2} \label{efe1p}\\
-{q\Lambda^0\over q-2}-{1\over 2}\;{ (n-2)(n-3)\over C^2}
&=&-\Lambda-  { (n-3)\over 2}\;\;
{\kappa^2 \eta^2\over C^2}\label{efe2p}
\eeqa

These can be solved for $\Lambda$ and $C^2$. We must consider
separately
the cases $\Lambda^0=0$ and $\Lambda^0 \neq 0$. In the latter case
\beq
\Lambda={q+n-3\over q-2}\;\Lambda^0,\q
C^2={[(n-2)-\kappa^2 \eta^2] (q+n-3) \over \Lambda}.
\label{sol}
\eeq

We see that for $q>2$, $\Lambda$ has the same sign as $\Lambda^0$. On
the
other hand, since $C^2>0$ it is clear that a solution with
a positive (negative) cosmological constant is consistent only
with a subcritical (supercritical) symmetry-breaking scale.
The critical value is $\eta^2_c\equiv (n-2)\kappa^{-2}$.
$n=2$ is an exception to this, since in this case the solution exists
only
for a negative cosmological constant.

When $\Lambda^0=0$, the solution only exists when $\eta=\eta_c$.
In this case
$\Lambda=0$  and
the value of $C^2$ is arbitrary.

We can summarize the above by saying that if we want to compactify
$n-1$ extra dimensions on an $(n-1)$-sphere with a global defect,
then we need $\eta>\eta_c$ if the background cosmological constant is
negative, $\eta=\eta_c$, when $\Lambda=0$ and
$\eta<\eta_c$
when the cosmological constant is positive.

The corresponding maximally symmetric geometries would be
$dS_q\times S^{n-1}$ when $\Lambda>0$, $M_q\times S^{n-1}$ 
when $\Lambda=0$ and
$AdS_q \times S^{n-1}$ when $\Lambda<0$ ($dS$ stands for deSitter, $M$ for
Minkowski and $AdS$ for anti-de Sitter). This compactification scheme is
in a way similar to that by Candelas and Weinberg \cite{cw}.
They consider gravity and a set of massless scalar fields.
The compactification arises from the quantum fluctuations of the fields
while here it arises from the monopole-like configurations of classical
fields, just like in \cite{Cremmer} or \cite{fr} where gauge monopoles 
were considered.

It may seem surprising that we can have static
solutions even in the case of vanishing cosmological constant. Since
the
curvature of the sphere acts as a cosmological constant in the
transverse
dimensions, one would not expect a static solution. However, we can see
from
the relations that this is possible when we introduce the
hedgehog configuration for the scalar fields.
In the transverse dimensions this also has the same effect as a
cosmological
constant and for the critical value of the symmetry-breaking scale its
value cancels exactly the contribution from the $(n-1)$-sphere curvature.

Although the above has been derived for a cosmological constant, the
result is more general. Consider now that
$G^0_{ij}=\kappa^2 T^0_{ij}$ with 
$T^0_{ij}=-\Lambda^0 g_{ij}+S^0_{ij}$ so that in the lower-dimensional 
energy-momentum tensor we have contributions from a cosmological constant
and some other form of energy which contributes through $S^0_{ij}$.
In higher dimensions we add the corresponding term to the 
energy-momentum tensor
\beqa
T_{ij}&\equiv& -\Lambda g_{ij}+S_{ij}
      -{1 \over 2}  (n-1)
{\eta^2 \over C^2} \;g_{ij},       \label{t1p}\\
T_{\alpha \beta}&\equiv&-\Lambda g_{\alpha \beta}+S_{\alpha \beta}
 -{1 \over 2}(n-3)
{\eta^2\over C^2} \;g_{\alpha \beta} \label{t2p}
\eeqa

This time we can find a solution to
EFE's if $\Lambda$ and $C^2$ solve equations
(\ref{efe1p}) and (\ref{efe2p}) as before, and 
if  $S_{ij}=S^0_{ij}$ and the angular components, $S_{\alpha \beta}$,
satisfy the relation
\beq
S_{\alpha \beta}=g_{\alpha
\beta} \bar{S}/(q-2) \q\q \mbox{with} \q\q
 \bar{S}\equiv g^{ij} S^0_{ij}.
\label{cond}
\eeq

Let's see that this is satisfied by massless scalar fields, which we
represent collectively as $\Psi$.
Our starting point is a
configuration that solves Einstein's equations in $q$ dimensions, so
the
fields are only functions of the coordinates $\{x^i\}$. In the
higher-dimensional solution with the global charge we maintain the same
ansatz and thus
\beqa
S_{ij}&=&-{1\over 2}\;g_{ij}\partial_k \Psi \partial^k \Psi
+\partial_i \Psi\partial_j \Psi \\
\bar{S}&=&-{q-2\over 2}\partial_k \Psi \partial^k \Psi \label{p1} \\
S_{\alpha \beta}&=&-{1\over 2}\;g_{ij}\partial_k \Psi \partial^k \Psi
\label{p2}
\eeqa

From equations (\ref{p1}) and (\ref{p2}) we can indeed see that the
relation $S_{\alpha \beta}=g_{\alpha \beta} \bar{S}/(q-2)$ is
satisfied.

One could also consider other sources, but then, this
compactification scheme would only work if supplemented by
the condition (\ref{cond}),
%$S_{\alpha \beta}=g_{\alpha \beta} \bar{S}/(q-2)$,
which
seems rather arbitrary. If we allow such arbitrariness,
we could consider the compactification of a
FRW-type solution. To construct the solution in $q+(n-1)$ dimensions
with $n-1>2$ compactified extra dimensions,
we start with a cosmological FRW solution with cosmological constant
in $q$ dimensions.
We can embed this solution into $n-1$ extra
dimensions if we consider an inhomogeneous fluid with
\beq
S^0_0=-\rho(t),  \q\q
S^{I}_J=\delta^I_J p(t), \q\q
S^{\alpha}_{\beta}=\delta^{\alpha}_{\beta} \bar{p}(t), \q\q
\mbox{with} \q \bar{p}(t)\equiv {(q-1)p(t)-\rho(t)\over q-2},
\label{emtp}
\eeq
where $\bar{p}$ is defined in such a way as to satisfy condition
(\ref{cond}).
Here we have split the $\{x^i\}$ coordinates as $\{x^0,x^I\}$,
the $0$ index referring to the time coordinate and
capital latin letters $I$ and $J$ to the spatial
coordinates. 

The metric would be
\beq
ds^2=-dt^2+a^2(t)\left\{ {dr^2 \over 1+kr^2}
+r^2 d\Omega^2_2 \right\}+C^2 d\Omega^2_{n-1}
\eeq
with $C^2$ given by (\ref{sol}).

By the same token,
one can also consider the braneworld solutions in codimension one.
In this models our universe is pictured as a domain wall and,
in a bulk based approach, solutions
can be obtained by considering a fixed background spacetime
and studying the moving wall trajectories \cite{ruth3}.
With
the procedure described previously we can get solutions
in which this domain wall has
$n-1>2$ compactified extra
dimensions. This gives explicit examples of solutions
in which although we have a number of extra dimensions greater that one,
only one can be considered large.

Considering for simplicity the $Z_2$ symmetric walls as an example,
we can write the metric for the background spacetime
\beq
ds^2=-h(r) dt^2+h^{-1}(r) dr^2+r^2\left[{d\chi^2  \over1-k \chi^2}
+\chi^2 d\Omega_2^2 \right]+C^2 d\Omega_{n-1}^2.
\eeq
See \cite{ruth3} for details on the form of $h(r)$.

The position of the wall can be parameterized by $\tau$, the proper time of an
observer comoving with the wall $r=R(\tau),\;\; t=T(\tau)$, where the
trajectory is ultimately determined by the matter on the wall.
For a given trajectory in the lower dimensional solution
there will be a corresponding
higher-dimensional one.
In the original solution there is a cosmological constant in the bulk and
a localized perfect fluid on the wall. In the higher-dimensional
generalization the bulk has an additional contribution from the scalar
fields and localized on the brane, which now has $n-1$ compactified extra
dimensions, we have an inhomogeneous perfect fluid as that of
equation (\ref{emtp}) in order to satisfy the condition
given by (\ref{cond}).

\section{Conclusions}
In this paper we have investigated the gravitational effect
of $n$ scalar fields with the simplest nontrivial topological
configuration, a radial configuration
such that $\vec{\phi}=\eta \hat{r}$, where $\hat{r}$ is the unit
vector on the $(n-1)$-sphere and $\eta$ the symmetry-breaking scale.

As it is well known, such configurations can give rise to
global defects. Here we have given the general form
of the gravitational field, far from the core,
of a global defect in the
hedgehog configuration in a $O(n)$ invariant solution.
We have found that
it amounts to the introduction of a deficit solid angle.
Thus we can easily add the gravitational field
of such a defect to any solution to Einstein's equations
with $O(n)$ invariance by
adding the corresponding solid angle deficit
which appears through an extra $1-(\kappa^2 \eta^2/(n-2))$
factor in the $O(n)$ invariant part of the metric.

We have also considered the case where there is no defect because
the nontrivial vacuum configuration can be maintained everywhere, as for
textures. We have seen that when $n$ scalar fields are
in the hedgehog configuration
this can lead to compactification of $n-1$ spatial dimensions.
In particular, we have
seen how the presence of the $n$ scalar fields can be used
to add $n-1$ compact spatial dimensions to any solution of Einstein's
equations with a cosmological constant and massless scalar
fields.

In the first case we have written down new solutions describing a
global
monopole in an expanding universe as well as black holes and black
$p$-branes with a global charge. In the second case we have given
explicit solutions where we add $n-1$ compact dimensions to known
braneworld solutions in wich our world is seen as a moving domain wall
in a static bulk.

\section{Acknowledgments}
I am grateful to Jose Juan Blanco-Pillado, Gia Dvali and
Alexander Vilenkin for helpful discussions.
This work was
supported by the Basque Government
under fellowship number BFI.99.89.

\end{document}